# Open Access Mandates and the 'Fair Dealing' Button

By Arthur Sale[1], Marc Couture[2], Eloy Rodrigues[3], Leslie Carr[4] and Stevan Harnad[5]

## Introduction and initial motivation

Many disciplines have a long history of distributing research findings by mail, even before the scholarly journal appeared. In a few fields, such as high energy physics and computer science, *preprints* used to be systematically mailed to a set of collaborating universities even before refereeing and publication (Goldschmidt-Clermont, L 2002; Postel & Reynolds 1985). In these and many other disciplines, however, researchers would mail a postcard to the author after publication to request a *reprint* of the published, refereed paper for research use. The author would then mail to the requester either a publisher-supplied reprint, or, if those ran out or were unavailable, a photocopy of the published article or of its refereed, revised and accepted final draft.

This practice was commonplace and accepted, explicitly or tacitly, by the journal publishers. It is at least half a century old (Bratt 1937; Garfield 1972).

Starting at least in the early 1980s, as the Internet began to transform the world of scientific and scholarly communication, researchers turned to using email in place of mail for the same purpose. The requests first still came by card, but in response *eprints* were sent by email, saving time, printing costs and postage. Soon journals began to list their authors' email addresses, to facilitate enquiries directed to the author. Scholarly research began to operate even more as a coordinated, collaborating community.

With the growth of the Internet in general, and of researchers' institutional home-pages in particular, and with it the mounting demand for free online access to refereed research, known as *Open Access* (OA), it became apparent that authors' Institutional Repositories (IRs) could make both requesting and providing eprints much easier, faster, surer and more efficient. The obvious and optimal option was for eprints to be deposited in their author's IR and immediately made OA, so anyone on the web could find and download them whenever they wished (Harnad 1995).

For articles published in the majority of journals this soon became possible in principle, because their publishers had endorsed this OA self-archiving by their authors immediately upon publication (SHERPA RoMEO). But the remaining journals either endorsed only the self-archiving of the unrefereed preprint, or imposed an embargo of 6–12 months or more before their authors could make their refereed final drafts OA, or they did not endorse OA self-archiving at all. Thus was born what would later be variously called the 'Request-a-copy'

[1] University of Tasmania, Hobart, Australia Arthur.Sale@utas.edu.au
[2] Université du Québec à Montréal (UQAM), Montréal, Canada couture.marc@teluq.uqam.ca
[3] Universidade do Minho, Braga, Portugal eloy@sdum.uminho.pt
[4] University of Southampton, Southampton, United Kingdom lac@ecs.soton.ac.uk
[5] Université du Québec à Montréal (UQAM), Montréal, Canada & University of Southampton, United Kingdom harnad@ecs.soton.ac.uk

Button, the 'Email Eprint Request' Button, the 'Fair Dealing' Button, the 'Fair Use' Button, etc.

The Button appears on an IR page describing the metadata for a refereed journal (or conference) article whose full-text is deposited in the IR as *Closed Access* rather than Open Access. The Button then makes it possible for would-be readers to request that the author email the eprint to them for individual research purposes under the provisions of fair dealing in the world's Copyright Acts.

## History of the development

The motivation for adding the Button to the EPrints Institutional Repository software (Tansley & Harnad 2000; Sponsler & Van de Velde 2001) was hence to provide authors with an alternative way to provide access on an individual request basis to papers that they had deposited in their IR as Closed Access rather than Open Access (Hitchcock 2006). In addition, the Button was conceived as a further incentive for institutions and funders to adopt mandates requiring IR deposit of all refereed journal articles.

This statement needs explanation. Although authors could already have been making at least 63% of their annual articles OA immediately upon publication with the publisher's approval, only 15% were actually being deposited. Institutions and funders accordingly began adopting deposit mandates (see ROARMAP; Sale 2006a,b,c,d), but because 37% of journals did not endorse immediate OA self-archiving, the mandates (known as *delayed-deposit mandates*) were weakened to allow deposit to be delayed as long as the publisher chose to embargo OA. Moreover, in the case of authors whose publishers could not be persuaded to accept an author addendum to the copyright agreement that would explicitly sanction OA self-archiving, these delayed-deposit mandates had to allow waivers whereby authors could opt out of depositing.

It is here that the Button is potentially useful and perhaps sees its greatest importance, for another kind of mandate is possible — the *immediate-deposit/optional-access* (IDOA) mandate (Harnad 2006; also called the "Dual Deposit-Release" Suber 2006). The idea is that once all refereed final drafts, without exception, are being deposited in the author's or fundee's IR immediately upon acceptance, whether in Closed Access or Open Access, the Button can immediately begin to allow users to request and authors to provide individual access to Closed Access papers to tide over research usage needs during an embargo period. All the OA deposits are of course open to the research community immediately.

The Button also allows institutions and funders to strengthen author-addendum/opt-out mandates (such as Harvard's; see ROARMAP) that require the author to negotiate re-use rights with the publisher. While it is true that an author addendum would retain authors' self-archiving rights as well as certain re-use rights, such mandates must always allow opt-out waivers by authors who are either unsuccessful in getting their publishers to adopt the author addendum or who do not elect to try. IDOA, in contrast, mandates immediate exception-free deposit regardless of whether the author opts out of negotiating the author addendum. Whatever its subsequent access fate, every born-digital object (the final accepted draft of an article) is captured by the institution in its records at the point in time that is the natural milestone in the author's workflow (the point of official acceptance for publication), which is also the optimal time to begin providing access (Swan & Carr 2008). The Button can then provide 'Almost Open Access' for Closed Access deposits under 'fair dealing'.

For these multiple reasons, the Button was developed for the EPrints software in 2006 by one of the present co-authors (LC; Hitchcock 2006) and then replicated in DSpace by another co-author (ER; Rodrigues 2006).

### How does the Button work?

As previously described, the Button appears alongside the metadata of articles deposited as Closed Access, either because the publisher has stipulated that there is to be an embargo period before deposits can be made OA, or because the publisher has stipulated that deposits are not to be made OA at all. Note that the following figures are simulated screenshots derived from a French-language Canadian repository running EPrints software (Archipel, at l'Université du Québec à Montréal).

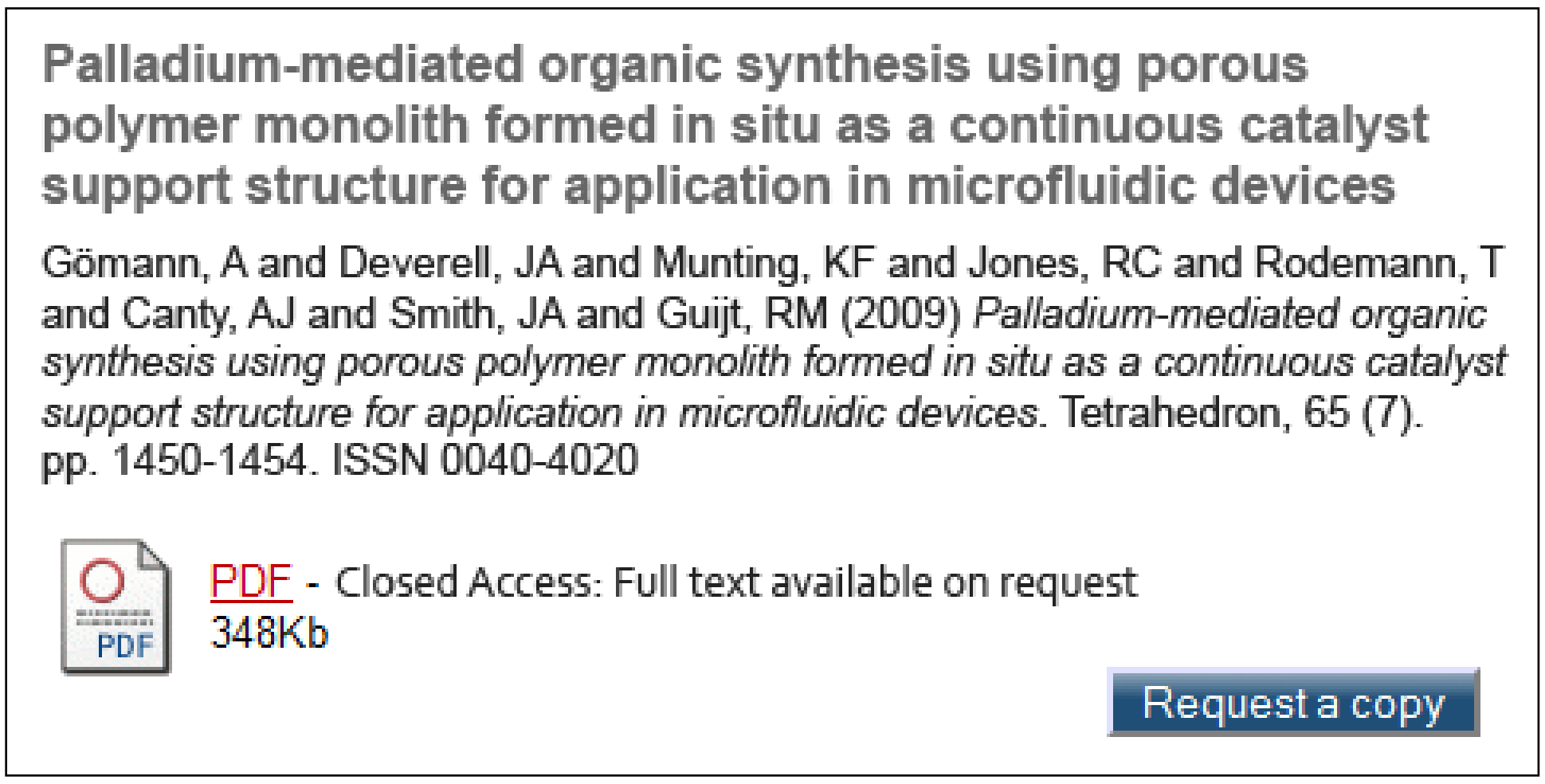

**Figure 1 — The Fair Dealing Button as the viewer sees it**

If users click once on the 'Request a copy' Button (action 1), they see a form page that asks them to paste in or type their email address (action 2) accompanied by a generic statement of the fair dealing conditions (in this example, according to Canadian copyright law, D'Agostino 2009) and a statement that the document is to be used according to these conditions. A click on a second 'Request a copy' Button (action 3) indicates assent and completes the transaction. Normally a following page acknowledges the request.

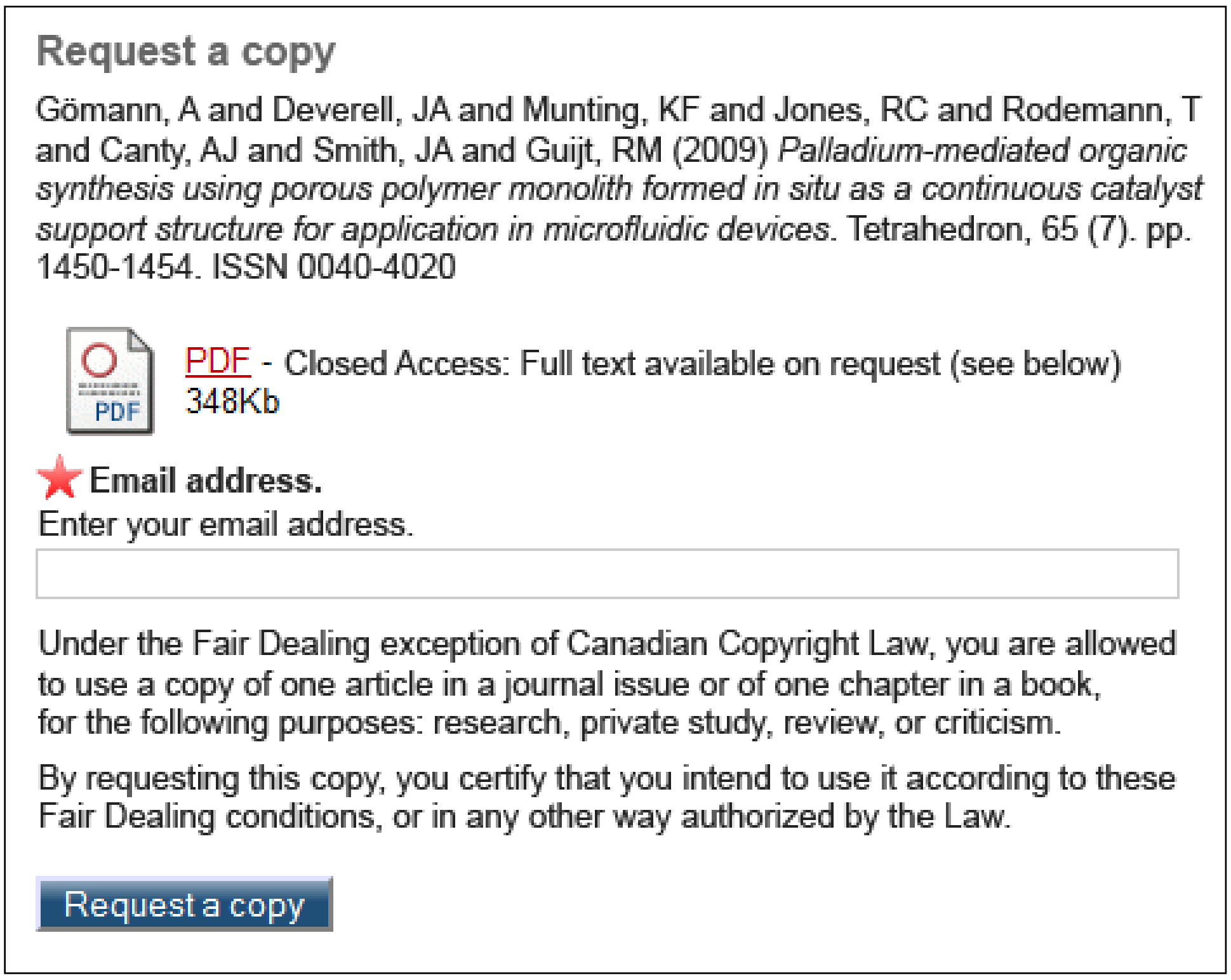

**Figure 2 — The request page as the viewer sees it**

Behind the scenes the repository looks up the depositing author of the article and sends an email requesting the author to authorize the sending of an electronic copy of the article to the requester. Note that the requester need never see or know the author's email address; nor does any crawler, thus reducing spam.

The author receives the email and is presented with two one-click alternatives: approve the request or deny it. The author can also choose to ignore the email. If the author approves the request, an electronic copy of the article is automatically emailed to the person who requested it. If denied, a short email is sent saying the request has been declined.

A few implementation details have been glossed over in the above description, such as how to identify the responsible author, what happens when the author has left the university, and what happens when the work is in two or more parts. These are minor problems for the repository software and management to handle, not the requester nor the approving author.

## Legal and policy considerations

Copyright is a real concern among repository managers as well as researchers, as has been found in earlier studies and regularly voiced by professors. Researchers are fearful of what they do not know, or do not understand. It is thus useful to clarify the legal status of the Button and its uses.

Closed Access (thus the Button) is usually chosen because the publisher, who has required the assignment of copyright or an exclusive license, has invoked consequential rights to forbid self-archiving or to impose an embargo period. The policies of such publishers normally imply that sending even a single copy of the article would be illegal unless explicitly authorized by the publisher.

However – and these policies usually don't mention it – the fair use (US) and fair dealing (Australia, Canada, UK, and other Commonwealth countries) provisions – or exceptions – of the copyright laws make this act legal if it satisfies certain criteria and, in the above-mentioned Commonwealth countries, if it is done for one of a few specific purposes, most of them common to all these jurisdictions: research, study, criticism, and news reporting.

What must be realized, however, is that in all jurisdictions court decisions involve a combination of many criteria, each of which leaves room for interpretation, their relative importance not precisely defined. In particular, legislators and courts have been reluctant to specify quantitative norms as to the amount of dealing that could maintain fairness. Australia is a notable exception: the Australian Copyright Act states that the reproduction, for research or study, of one article in a journal issue or one chapter in a book will automatically be deemed fair dealing.

The result is a case-by-case approach that makes difficult, except maybe in Australia, any firm prediction about a situation not identical or highly similar to one brought before the courts previously. Fortunately, in Canada, whose fair use provision in the law is by far the most succinct, the main reference for copyright issues relating to fair dealing is the 2004 Supreme Court judgment *CCH Canadian Ltd*. v. *Law Society of Upper Canada*, in a case where an access procedure very similar to the Button was involved. The case concerned a Bar Association library that was sued by publishers for, amongst other activities, sending by fax upon request from "lawyer[s], law student[s], member[s] of the judiciary or authorized researcher[s]", photocopies of various documents, including articles. In particular, the judgment (which was completely favourable to the Bar Association) states:

> [...] the Great Library will typically honour requests for a copy of one case, one article or one statutory reference. [...] This suggests that the Law Society's dealings with the publishers' works are fair. [...] the dealings might not be fair if a specific patron of the Great Library submitted numerous requests for multiple reported judicial decisions from the same reported series over a short period of time... (¶68)
>
> [...] a series of repeated fax transmissions of the same work to numerous different recipients might constitute communication to the public in infringement of copyright. (¶78)

Replacing 'fax' by 'email', which is a very small step, one sees that the reported scarcity of use of the Button compared to direct downloading of Open-Access documents (see below), would be a strong argument in favour of fair dealing in case of a challenge. Arguably, there could be infringement only in two cases: many users requesting the same article in a short period, and a user requesting many articles published in the same journal issue, or many chapters in a single book (for instance, a collective work).

An interesting statement, in the same Canadian Supreme Court judgment, is that "research must be given a large and liberal interpretation in order to ensure that users' rights are not unduly constrained, and is not limited to non-commercial or private contexts."

Another relevant part of the judgment concerns the Bar Association Access Policy, which played an important role in the Court's decision. The policy enumerates the types of works and portion thereof that can be sent to requesters, and provides for a manual review process of potentially infringing requests. As the Court concluded:

> The [library] Access Policy and its safeguards weigh in favour of finding that the dealings were fair. It specifies that individuals requesting copies must identify the purpose of the request for these requests to be honoured, and provides that concerns that a request is not for one of the legitimate purposes under the fair dealing exceptions in the Copyright Act are referred to the Reference Librarian. This policy provides reasonable safeguards that the materials are being used for the purpose of research and private study.(¶66)

The first part of this policy suggests that it would be advisable for institutional repository managers to (1) display clear statements on the IR website about what should – and shouldn't – be downloaded, both in the case of Open- and Closed-Access; (2) put explanations in the e-mail generating page associated with the Button and in the e-mails themselves about what is at risk of being considered unfair.

From: Archipel [mailto:archipel-admin@uqam.ca]

Sent: 28 Jan. 2007 16:18

To: Gömann, Anissa

Subject : Archipel: Request for "Palladium-mediated organic synthesis using porous polymer monolith formed in situ as a continuous catalyst support structure for application in microfluidic devices"

This document has been requested by requester@someplace.ca for the purpose of research, private study, criticism or news reporting, or for another use allowed by the Law. Please can you respond.

Gömann, Anissa et al.(2009). Palladium-mediated organic synthesis using porous polymer monolith formed in situ as a continuous catalyst support structure for application in microfluidic devices. *Tetrahedron*, 65(7): 1450-1454.

<http://www.archipel.uqam.ca/00/>

Note. Accepting a large number of requests for the same document in a short period, or requests for more than one article in the same journal issue or more than one chapter in a book may result in copyright infringement.

Click here to send the requested document.
<http://www.archipel.uqam.ca/perl/users/respond_doc?eprintid=00&email=requester@someplace.ca&action=accept>

Click here to reject the request.
<http://www.archipel.uqam.ca/perl/users/respond_doc?eprintid=00&email=requester@someplace.ca&action=reject>

Archipel
http://www.archipel.uqam.ca
archipel-admin@uqam.ca

**Figure 3 — The authorization email sent to the author following a user request.**

As to the second part of the policy, one could envision an automated monitoring of potentially unfair dealings. For example, one could try to detect an abnormally high rate of (accepted) requests for a single article. One could also try to detect successive requests, by the same user, of many articles in the same journal issue or many chapters in the same book, though this would be more difficult, in part because the articles or chapters could have been deposited individually by their respective authors in different repositories.

In both cases, the goal would not be to block these possibly infringing uses, but to inform or remind the author that acceptance of such requests could constitute an infringement of copyright. This is one lesson from the Canadian Supreme Court judgment, which overthrew the lower Court of Appeal decision in which fair dealing was not deemed proven because no one could guarantee that all the material provided was used in a 'fair dealing manner'. As the Supreme Court explains:

> [...] is it incumbent on the Law Society to adduce evidence that every patron uses the material provided for in a fair dealing manner or can the Law Society rely on its general practice to establish fair dealing? I conclude that the latter suffices.

## The IDOA mandate

The Button, as has been seen, derived from the fair dealing provisions of the various Copyright Acts around the world, and longstanding publisher acceptance. It extends the capability of acquiring a copy of the Closed Access deposit even to researchers in disciplines where reprint requests were rare. Moreover, although in the NISO terminology the deposit is just the 'Accepted Manuscript' rather than the 'Version of Record' (VoR), this is a minor difference for users who would otherwise have no access at all. The bibliographic metadata of the canonical VoR are in any case available if the work needs to be cited.

The most important factor driving the implementation and spread of the Button is the IDOA mandate. Previously, making the world's refereed research accessible to the world had been focused on trying to persuade publishers to approve self-archiving of the author's Accepted Manuscript (otherwise called the final draft). It was assumed that deposit could only be mandated in those cases which the publisher had formally approved, because deposit was conflated with OA access-setting.

With the Button, deposit is separated from access-setting, and the mandate only governs deposit and its timing. Institutions can require their authors to deposit all final drafts, without exception. The Button makes depositing Closed Access papers worthwhile by allowing authors to provide 'Almost OA' to them on an individual request by request basis. For the would-be reader, almost-OA is still free but involves a delay in receiving the eprint requested, and a risk that it might not be provided. IDOA plus the Button make the universal adoption of deposit mandates both legal and useful.

It is hard to overstate the potential importance of this simple practical and technical development. Mandatory deposit of *all* Accepted Manuscripts can now be required by *all* universities, research centers, and grant-giving bodies worldwide, without exception. This covers all of the OA movement's target content.  However, the full implications of this new possibility have not yet been fully realized let alone exploited. Only 101 institutional and departmental mandates and 42 funder mandates have been registered with the global mandate directory ROARMAP at the time of writing.

## The mandatory background

In the Web era, the optimal way for authors to make eprints of their published, peer-reviewed journal articles accessible to would-be users is simply to deposit them in their institutional repositories, freely accessible to everyone, webwide, any time on demand.

For the authors of articles published in the 37% of journals that have either not endorsed making such OA deposits, or have stipulated an embargo of 6-12 months or longer, IDOA still makes it possible to capture the author's born-digital Accepted Manuscript at the time of acceptance, the natural point in the author's work-cycle at which to deposit a new paper. A variety of options then open up.

First, it is now possible to enter an embargo expiry date in the repository at deposit time; once the expiry date is reached the article automatically becomes OA, without further intervention from the author or the repository staff. This is efficient for both the author and the repository manager.

Second, the Button enables researchers worldwide to request the Closed Access articles too. This 'Almost OA ' preserves the central aim that the world's research should be open to any researcher with Internet access, free of charge. What it lacks is immediacy and certainty, in that the copy is not delivered immediately, but after a delay which might vary from minutes to a day or two depending on time zones, weekends, etc, and there is also the possibility that an author might ignore the request, not approve it for some reason, or under some circumstances not receive it.

For these reasons, use of the Fair Dealing Button cannot be regarded as full Open Access, but it comes close.

The terms of existing OA mandates vary significantly. Some mandates specify deposit in a subject repository, some in an institutional repository. Some insist that their contract with the researcher predates and limits any contracts that the researcher may make with a publisher and hence over-rides all such later contracts. Some offer funding for upfront open access fees where they exist (for example some Open Access journals).

The momentum for OA has altered the attitude and tactics of publishers: realizing that progress towards OA is unstoppable, many publishers have endorsed 'Green' OA self-archiving of the Accepted Manuscript by their authors. Some converted to OA publishing, making their own online Version-of-Record open access. Some also began to offer hybrid models. In a hybrid model, the journal is still toll-restricted to subscribers during an embargo period, but authors can buy immediate open access rights for their individual article for a fee. Regardless of the price and value of such a purchase, it illustrates that the scholarly publishing industry is in the middle of inexorable change and is adapting to the challenge of the Internet.

Of course, not all research is funded by grants, but across all fields virtually all research is produced by universities or research centers. This is why institutional mandates are even more important than funder mandates. Moreover, the optimal locus of deposit for both kinds of mandates is the author's own institutional IR. The number of research institutions in the world is large but finite and changes occur slowly. The IRs are all OAI-compliant, and hence are interoperable. This means that harvesting and aggregating of IR metadata for seamless joint search and retrieval is feasible, as if it were all in one global repository; several examples of such central harvesters exist at national level (Australian Research Online) and global level (Citeseer$^{x}$, Base and OAIster).

It is also possible to create regional custom search engines based on institutional repositories, building on the indexes created by major search engines such as Google. For example, one author has created a custom search engine for Australia and New Zealand (AuseSearch), and another for Africa (eSearch-Africa).

## Take-up and usage

Research on the Fair Dealing Button is only just commencing since it has not been in use for long. In what follows it should be noted that common search engines return an OA full-text page as a higher ranking item than its metadata page, so most views of Open Access items go to that page. Only if the item is Closed Access will the metadata page appear near the top of the ranking, where the viewer can see the Button. However, in gateways like Base and OAIster which harvest only metadata, the metadata page will be presented as the norm and the only version.

The following data were provided from the following IDOA mandated institutions:

- University of Southampton,
- University of Stirling, and
- Universidade do Minho.

| Author responses | University of Southampton[a] (UK) | University of Stirling[b] (UK) | Universidade do Minho[c] (Portugal) |
|---|---|---|---|
| Approved | 47 % | 60 % | 27 % |
| Ignored / unanswered | 53 % | 37 % | 72 % |
| Rejected / denied | < 1 % | 3% | 1 % |

a. Aug. 2008 – Jan. 2010; b. Apr. 2009 – Jan. 2010; c. Jan. – Dec. 2009

**Figure 4 — Responses to requests in three university-based repositories**

The approval success rate varies from 27% to 60%. Very few requests are actively denied. The majority of unapproved requests are probably due to non-receipt of the email or uncertainty regarding the legal status of the request; some repositories report author fatigue of dealing with requests being a factor.

## Putting the Button into perspective

Given a significant number of Button requests which are ignored or lost, one might be tempted to assume that it has not worked. However, this is not true. The principal impact of the Button has been to enable the adoption of institutional IDOA mandates. Deposit is mandated immediately without legal constraints, with the Button serving to assist authors interested in the dissemination of their articles. To put this into context, data from two universities with long-standing mandates are shown in Fig. 5. It shows that the Button applies to 5-7% of the deposited articles, but without it, all the other open access articles might be missing. The variation in the proportion of Closed Access articles is being researched.

| Number of articles | University of Southampton[a] (UK) | Universidade do Minho (Portugal) |
|---|---|---|
| Total | 7 864 | 7 515 |
| Closed Access | 551 (7 %) | 353 (5 %) |

a. 2001-date

**Figure 5 — Closed Access in two university repositories**

## Summary

The Fair Dealing Button has facilitated the adoption of IDOA mandates around the world and in this it has had a sizeable impact. The growth in these mandates creates the climate for universal OA to the world's research for all researchers, which is what all researchers want for their own publications so they can maximise the uptake, usage and impact of their work. That is why they give the rights in their articles away for free in the first place (Lawrence 2001, Swan 2006; Harnad et al 2008, 2009; Hitchcock 2010).

Even more importantly, IDOA mandates coupled with the Button take the publisher completely out of the loop insofar as the adoption of and compliance with a deposit mandate is concerned: publishers have no say over whether or when a deposit is made in an institutional database; they only have a say in whether or when the deposit is made Open Access or Closed Access, possibly embargoed until a designated date.

The IDOA mandate and the Button divide and conquer. All hesitations about whether and when a university or research center (or funder) can mandate deposit itself are rendered irrelevant. The Button provides 'Almost OA' for embargoed content during the embargo, and also for permanently Closed Access content. Immediate deposit can be universally mandated by all funders, all universities and all research centers: there is no remaining need to worry about the legality of adopting a mandate at all, nor any need to allow opt-outs, waivers or delayed deposit, because obligatory deposit is separated from the optional OA-setting, with the Fair Dealing Button bridging the gap for those who cannot provide OA immediately.

Researchers from all disciplines can be confident that the couple of clicks required to give a fellow researcher access to their Closed Access article is legal... and fair.